\documentclass[apjl]{emulateapj}

\usepackage{amssymb}
\usepackage{graphicx,graphics}% Include figure files
\usepackage{epsfig}% Include figure files
\usepackage{dcolumn}% Align table columns on decimal point
\usepackage{bm}% bold math
\usepackage{natbib}
\usepackage{times}
\usepackage{ulem}
\newcommand{\ci}{C{\sc i}}
\newcommand{\zdla}{z_{\rm DLA}}
\newcommand{\zqso}{z_{\rm QSO}}
\newcommand{\tdv}{\int \tau {\rm dV}}
\newcommand{\ee}{\times 10^{21}}
\newcommand{\sitwostar}{S{\sc ii}*}

\newcommand{\mgtwo}{Mg{\sc ii}~$\lambda$2796}

\newcommand{\kms}{km\,s$^{-1}$} 
\newcommand{\ts}{\ensuremath{{\rm T_s}}}

\newcommand{\nhi}{\ensuremath{\rm N_{HI}}}

\newcommand{\cm}{cm$^{-2}$}
\newcommand{\hi}{H{\sc i}}
\newcommand{\hii}{H{\sc i} 21cm}

\shorttitle{GMRT detections of H{\sc i} 21cm absorption at $z \approx 2$}
\shortauthors{Kanekar et al.}
\begin{document}
\title{Giant Metrewave Radio Telescope detection of two new H{\sc i} 21cm 
absorbers at $z \approx 2$}

\author{N. Kanekar\altaffilmark{1},
}
\altaffiltext{1}{Ramanujan Fellow, National Centre for Radio Astrophysics, 
TIFR, Ganeshkhind, Pune - 411007, India; nkanekar@ncra.tifr.res.in}

\begin{abstract}
I report the detection of H{\sc i}~21cm absorption in two high column density 
damped Lyman-$\alpha$ absorbers (DLAs) at $z \approx 2$ using new wide-band 
$250-500$~MHz receivers onboard the Giant Metrewave Radio Telescope. The integrated H{\sc i}~21cm 
optical depths are $0.85 \pm 0.16$~km~s$^{-1}$ (TXS1755+578) and $2.95 \pm 0.15$~km~s$^{-1}$
(TXS1850+402). For the $z=1.9698$ DLA towards TXS1755+578, the difference in H{\sc i}~21cm 
and C{\sc i} profiles and the weakness of the radio core suggest that the H{\sc i}~21cm 
absorption arises towards radio components in the jet, and that the optical and radio 
sightlines are not the same. This precludes an estimate of the DLA spin temperature. For 
the $z = 1.9888$ DLA towards TXS1850+402, the absorber covering factor is likely to be 
close to unity, as the background source is extremely compact, with all the 5~GHz emission 
arising from a region of size $\leq 1.4$~mas. This yields a DLA spin temperature of ${\rm T_s}
= (372 \pm 18) \times (f/1.0)$~K, lower than typical ${\rm T_s}$ values in high-$z$ 
DLAs. This low spin temperature and the relatively high metallicity of the $z = 1.9888$
DLA ([Zn/H]~$= (-0.68 \pm 0.04)$) are consistent with the anti-correlation between 
metallicity and spin temperature that has been earlier found in damped Lyman-$\alpha$ 
systems.

\end{abstract}


\maketitle
\section{Introduction} 
\label{sec:intro}

\hii\ absorption studies have long been used to study neutral hydrogen (\hi) in both 
galaxies fortuitously located along the line of sight to background radio-loud quasars
and gas associated with active galactic nuclei. In the case of absorption-selected galaxies at 
high redshifts, the damped Lyman-$\alpha$ absorbers \citep[DLAs;][]{wolfe05}, the \hii\ optical 
depth can be combined with the \hi\ column density (inferred from the Lyman-$\alpha$ absorption profile) 
to yield the spin temperature of the neutral gas along the sightline \citep[e.g.][]{kanekar04}. 
\hii\ absorption studies thus provide one of the few direct probes of physical conditions in the 
neutral gas in high-redshift DLAs \citep[e.g.][]{carilli96,chengalur00,kanekar03}, complementing 
optical and ultraviolet estimates of elemental abundances, molecular fractions, dust depletions, kinematic 
widths, etc, from observations of low-ionization metal and molecular hydrogen lines 
\citep[e.g.][]{pettini94,prochaska03a,noterdaeme08,rafelski12}.

Early \hii\ absorption studies of high-redshift DLAs suggested that conditions in the 
neutral interstellar medium in these galaxies were different from those in the Milky Way.
The few high-$z$ DLAs with \hii\ absorption studies were found to have relatively high 
spin temperatures ($\ts \gtrsim 1000$~K), nearly five times higher than typical values
in the Milky Way \citep[e.g.][]{dickey78,wolfe79,wolfe82,wolfe85,braun92,carilli96,debruyn96,briggs97,kanekar97}. 
Unfortunately, the poor low-frequency coverage of radio telescopes meant that such 
studies were only possible in a handful of DLAs at $z \gtrsim 2$. To make matters worse,
few DLAs were known at low redshifts, due to which there were also hardly any 
$\ts$ measurements in DLAs at $z < 1$.

The situation has changed significantly over the last decade, with the advent of the Green 
Bank Telescope (GBT) and the Giant Metrewave Radio Telescope (GMRT) which combine high 
sensitivity with excellent low-frequency radio coverage. There are now more than 50~DLAs 
at all redshifts with \hii\ absorption studies, nearly 40~of which have estimates of the 
spin temperature after correcting for the absorber covering factor 
\citep[see][and references therein]{kanekar14}. Targeted Lyman-$\alpha$ spectroscopy of 
strong \mgtwo\ absorbers \citep[][]{rao00,rao06} 
and radio-loud quasars \citep[][]{ellison01,ellison08,jorgenson06} have yielded new samples 
of DLAs at all redshifts for follow-up \hii\ absorption studies. Conversely, more than 
twenty-five redshifted \hii\ absorbers at $z < 1.7$ have been identified from direct spectroscopy 
of strong \mgtwo\ absorbers \citep[e.g.][]{lane98,lane02,kanekar09b,gupta07,gupta09,gupta12}.
Follow-up Lyman-$\alpha$ spectroscopy of some of these systems has provided estimates of the 
\hi\ column density, and hence, of the absorber spin temperature \citep[][]{ellison12}.
Finally, high-resolution optical spectroscopy of systems with \hii\ absorption studies 
have been used to infer their gas-phase metallicities, dust depletions, etc, for comparison
with the measured spin temperatures \citep[e.g.][]{kanekar09c,kanekar14,srianand12,ellison12}.

Based on the above studies, it is now apparent that the spin temperature distribution 
in DLAs is significantly different from that in the Milky Way, with DLAs having typically 
higher spin temperatures \citep[][]{kanekar03,kanekar14}. The anti-correlation detected 
between spin temperatures and metallicities in DLAs indicates that their high $\ts$ values 
arise due to larger fractions of the warm phase of neutral hydrogen in the absorbers, 
probably due to fewer radiative cooling routes at their typically-low metallicities 
\citep[][]{kanekar01a,kanekar09c,kanekar14}. DLA spin temperatures also show redshift 
evolution: absorbers at $z \gtrsim 2.4$ have both fewer detections of \hii\ absorption and 
higher spin temperatures than systems at $z \lesssim 2.4$, again due to their smaller cold gas 
fractions \citep[][]{carilli96,chengalur00,kanekar03,kanekar14}. 

Despite the recent progress in the field, an important lacuna remains the paucity of 
{\it detections} of \hii\ absorption at high redshifts, $z \gtrsim 2$. There are, at 
present, only seven detections of \hii\ absorption in DLAs at $z \gtrsim 2$ 
\citep[][]{wolfe82,wolfe85,kanekar06,kanekar07,kanekar13,york07,srianand12}. While 
\hii\ non-detections provide lower limits to $\ts$, which are useful in testing its 
redshift evolution, a detailed understanding of the evolution of the spin temperature 
and its relation with quantities like metallicity, dust depletion and star formation rate, 
as well as modeling of local physical conditions in the gas, critically require 
detections of \hii\ absorption. \hii\ line detections at high redshifts are also needed 
to use the lines in conjunction with ultraviolet resonance lines to probe the possibility 
of fundamental constant evolution \citep[][]{wolfe76b,kanekar10,rahmani12}. Until 
now, the GBT has been the only telescope providing wide frequency coverage below 1~GHz;
unfortunately, the GBT is a single dish and hence more susceptible to the terrestrial
interference that is common at these low radio frequencies. I report here two new detections 
of redshifted \hii\ absorption at $z \approx 2$, first results from new wide-band 
$250-500$~MHz receivers that are currently being installed on the GMRT.

\section{Observations, data analysis and spectra}
\label{sec:data}

The GMRT is presently being upgraded with a new suite of receivers that will give 
the telescope near-seamless coverage between 120~MHz and 1437~MHz. As part of 
this upgrade, the present 327~MHz receivers (which covers $\approx 295-365$~MHz)
are being replaced by wide-band cone-dipole receivers covering $\approx 250-500$~MHz,
and with higher sensitivity than that of the earlier P-band system \citep[][]{bandari13}. 
The initial
commissioning tests of the new receivers yielded tentative detections of redshifted
\hii\ absorption in two DLAs at $z = 1.9698$ towards TXS1755+578 and $z = 1.9888$ 
towards TXS1850+402 \citep[][]{jorgenson06}. These two DLAs were hence re-observed
in July~2013 to confirm the detections of \hii\ absorption.

The GMRT observations of the two $z \approx 2$ DLAs were carried out on 2013, July~31, 
using the eleven antennas equipped with the new cone-dipole receivers, and the GMRT Software 
Backend. Bandwidths of 1.04~MHz (TXS1850+402) and 4.17~MHz (TXS1755+578) were used for the 
observations, sub-divided into 512 channels, and centred at the redshifted \hii\ line 
frequency (478.28~MHz for TXS1755+578 and 475.24~MHz for TXS1850+402). Observations of 3C286 
and 3C48 were used to calibrate the flux density scale, and of 3C380 to calibrate the 
antenna bandpasses and initial antenna gains. The total on-source time was 3.5~hours for 
each source.

The GMRT data were analysed in ``classic'' {\sc aips}, using standard procedures. After initial 
data editing and calibration of the antenna gains and bandpasses, about 50~channels on each
target source were averaged into a ``channel-0'' dataset. A standard self-calibration procedure 
was then used to obtain the antenna gains, with a few rounds of phase-only self-calibration and 
3-D imaging, followed by amplitude-and-phase self-calibration, 3-D imaging and data editing. This 
procedure was continued until the image did not improve on further self-calibration, and no 
evidence was found for bad data. In both cases, the target source was the strongest source in 
the field, with flux densities (measured using {\sc jmfit}) of $377.4 \pm 1.3$~mJy (TXS1755+578)
and $650.3 \pm 3.2$~mJy (TXS1850+402). Neither source showed evidence of extended emission in
the GMRT images. The final image was then subtracted from the calibrated spectral-line 
visibilities using the task {\sc uvsub}, and {\sc uvlin} then used to subtract any residual 
emission via a linear fit to each visibility spectrum. Finally, {\sc cvel} was used to shift the 
residual visibilities to the heliocentric frame. The data were then imaged and a spectrum obtained 
by taking a cut through the spectral cube at the location of the target sources. 

\begin{figure*}[t!]
\centering
\includegraphics[scale=0.4]{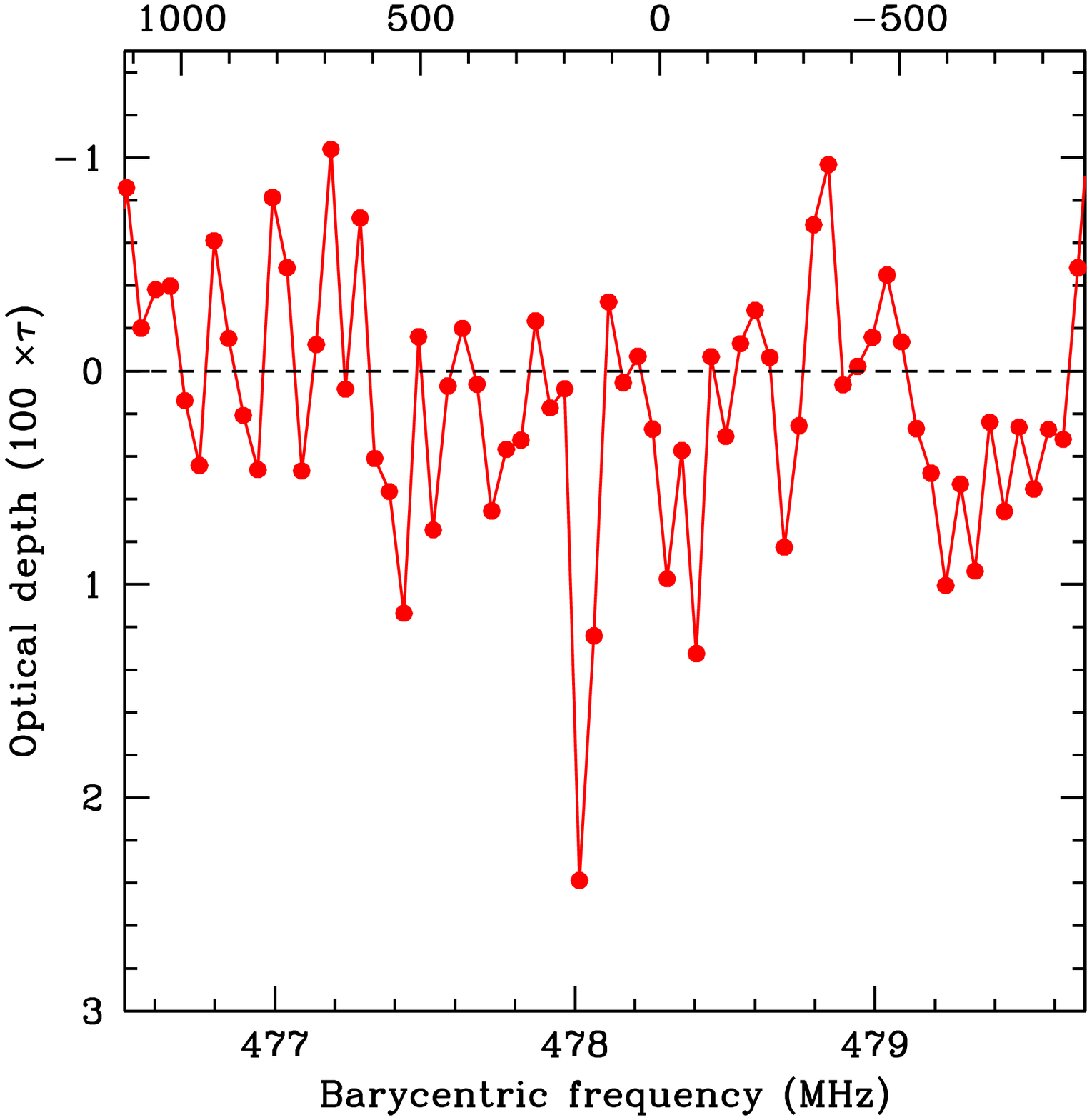}
\includegraphics[scale=0.4]{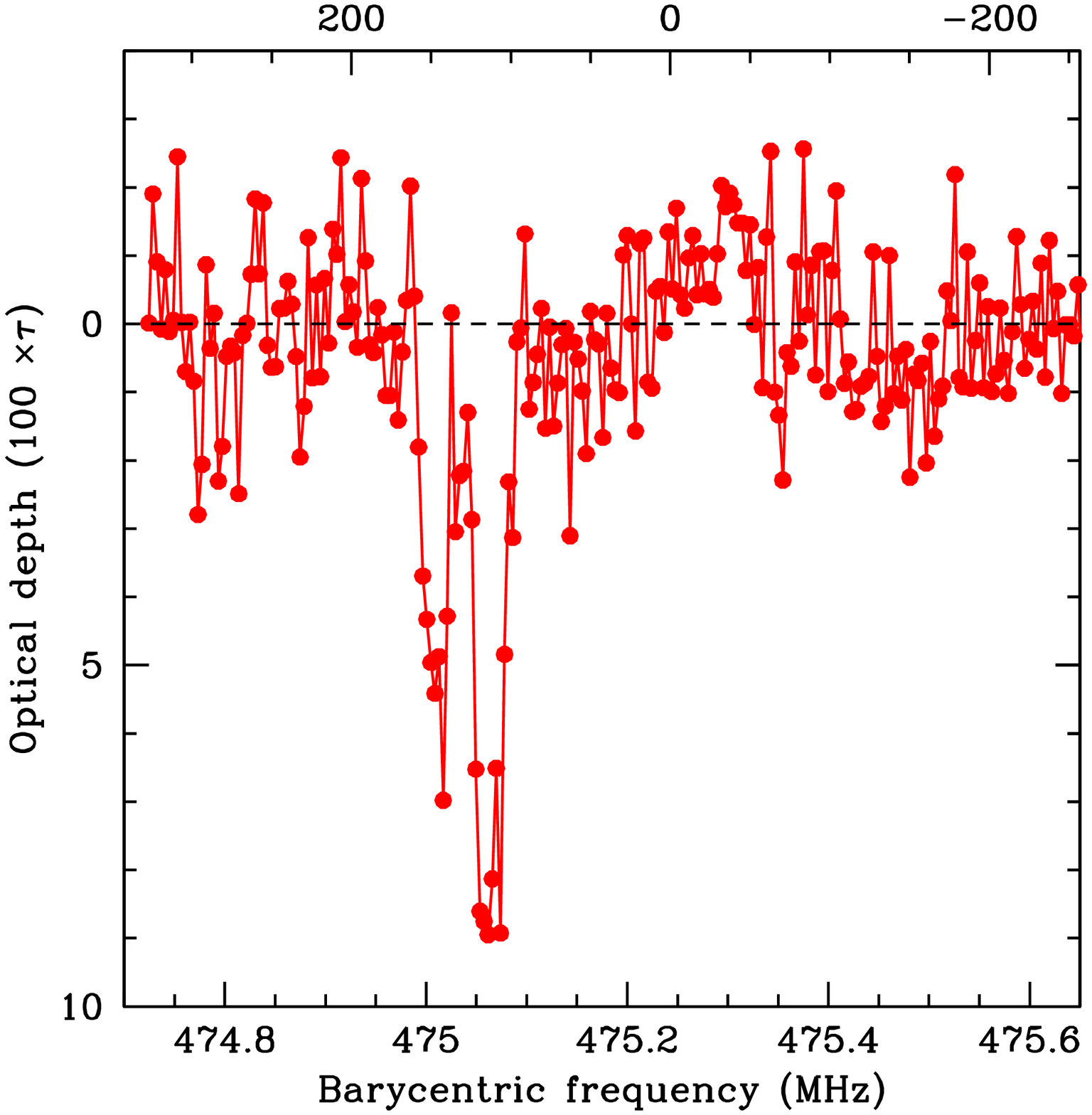}
\caption{GMRT \hi\ absorption spectra from [A]~the $z = 1.9698$ DLA towards TXS1755+578
(left panel) and [B]~the $z = 1.9888$ DLA towards TXS1850+402 (right panel). In both 
panels, \hii\ optical depth ($100 \times \tau_{\rm 21cm}$) is plotted against heliocentric
frequency, in MHz. The top axis of both panels shows velocity, in \kms, relative to 
$z = 1.9698$ (left panel) and $z = 1.9888$ (right panel). In the case of TXS1755+578, the 
spectrum has been smoothed to, and resampled at, a velocity resolution of $\approx 31$~\kms.}
\label{fig:spc}
\end{figure*}

The final GMRT spectra towards TXS1755+578 and TXS1850+402 are shown in the two panels 
of Fig.~\ref{fig:spc}, with optical depth plotted versus heliocentric frequency. The spectrum 
towards TXS1850+402 has been Hanning-smoothed and re-sampled, and has a velocity resolution 
of $\approx 2.5$~\kms, while that towards TXS1755+578 has been further boxcar-smoothed by 
3~channels and re-sampled, and has a velocity resolution of $\approx 31$~\kms. 
The root-mean-square (RMS) optical depth noise values on the original Hanning-smoothed and re-sampled 
spectra are 0.011 per 4.1~kHz channel (TXS1850+402) and 0.0092 per 16.3~kHz channel (TXS1755+578); 
note that significantly more data were edited out for the latter source, due to intermittent 
radio frequency interference. Both spectra show evidence of \hii\ absorption, with integrated 
\hii\ optical depths of $2.95 \pm 0.14$~\kms\ (TXS1850+402) and $0.85 \pm 0.16$~\kms\
(TXS1755+578). Note that the \hii\ absorption towards TXS1755+578 is relatively weak, 
and extends across only two independent $31$~\kms\ channels. However, the feature has $> 5\sigma$ 
significance and was detected in both observing runs, separated by many months. It is hence 
likely to be real.

\section{Results and Discussion}
\label{sec:discussion}

\setcounter{table}{0}
\begin{table*}
\begin{center}
\caption{Results
\label{table:results}}
\begin{tabular}{|c|c|c|c|c|c|c|c|c|}
\hline
             &         &        &       &   &   &  &       &   \\
QSO          & $\zqso^A$ & $\zdla^A$ & $N_{\rm HI}^A$     &      $S_\nu^B$    & $\tdv$         & $\Delta V_{\rm 90}^C$ & $f^D$   & $\ts^E$  \\
             &         &         & $\ee$~\cm\       &       mJy       &   \kms\         &  \kms               &       &   K    \\
\hline
             &         &         &        &       &   &  &       &   \\
1755+578  & $2.110$ & $1.9698$ & $2.51 \pm 0.15$ & $377.4 \pm 1.3$ & $0.85 \pm 0.16$ & $41$                & $0.15$ & $-$    \\
1850+402  & $2.120$ & $1.9888$ & $2.00 \pm 0.25$ & $650.3 \pm 3.2$ & $2.95 \pm 0.14$ & $100$               & $1.0$  & $(372 \pm 18) $ \\
             &         &         &        &   &   &  &       &   \\
\hline
\end{tabular}
\vskip 0.05in
Notes: (A)~The values of $\zqso$, $\zdla$ and $\nhi$ are from \citet{jorgenson06}.\\
(B)~The flux density $S_\nu$ is at the redshifted \hii\ observing frequency. \\
(C)~$\Delta V_{\rm 90}$ is the velocity range containing 90\% of the integrated 
\hii\ optical depth \citep[e.g.][]{prochaska97}. \\
(D)~The covering factor $f$ has been estimated from high-frequency VLBI studies; 
see main text for details. \\
(E)~The DLA spin temperature $\ts$ is not listed for the $z = 1.9698$ 
DLA towards TXS1755+578 as it appears plausible that the radio and optical sightlines 
are not the same for this source; see main text for details.
\end{center}
\end{table*}

For \hii\ absorption against a compact radio source, the \hi\ column density ($\nhi$), the 
\hii\ optical depth ($\tau_{\rm 21cm}$) and the spin temperature ($\ts$) are related by the 
equation \citep[e.g.][]{rohlfs06}
\begin{equation}
\nhi = 1.823 \times 10^{18} (\ts/f) \int \tau_{\rm 21cm} {\rm dV} \:,
\label{eqn:tspin}
\end{equation}
\noindent where $f$ is the absorber covering factor, giving the fraction of the background radio
emission that is occulted by the foreground DLA. The covering factor can be estimated from 
very long baseline interferometry (VLBI) observations at or near the redshifted \hii\ line 
frequency to determine the fraction of flux density in the compact radio core 
\citep[e.g.][]{briggs83,kanekar09a,kanekar14}. Note that a critical assumption in the above 
equation is that the \hi\ column density measured along the optical sightline is 
the same as that along the radio sightline. The \hi\ column densities of the DLAs towards 
TXS1755+578 and TXS1850+402 are $(2.51 \pm 0.15) \times 10^{21}$ and $(2.00 \pm 0.25) \times 10^21$, 
respectively. Equation~\ref{eqn:tspin} then yields $\ts = (1612 \pm 305) \times f$ (TXS1755+578) 
and $\ts = (372 \pm 18) \times f$~K (TXS1850+402). The above results are summarized in 
Table~\ref{table:results}.

Unfortunately, there are at present, no low-frequency VLBI observations of TXS1755+578 and 
TXS1850+402 from which one might directly measure the fraction of flux density in the radio
core. However, the two sources have either inverted (TXS1755+578) or flat (TXS1850+402) spectra
at low frequencies; such spectra typically arise due to synchrotron self-absorption, 
indicating that the radio emission is very compact. One hence would expect a relatively 
high core fraction in both sources. In the case of TXS1850+402, VLBI observations have
found all the 5~GHz emission to arise from two components both lying in a region smaller than 
$\approx 1.4$~mas \citep[][]{henstock95,pollack03}. In combination with the flat spectrum of 
TXS1850+402, this suggests that the covering fraction is likely to be close to unity at low 
frequencies. Conversely, 5~GHz VLBI studies of TXS1755+578 have shown that the source has 
multiple components extended over 30~mas, with a core flux density of $\approx 58$~mJy and a 
total flux density of $\approx 396$~mJy in the VLBI image \citep[][]{pollack03,helmboldt07};
this suggests a core fraction of $\approx 0.15$. Including these in the spin temperature estimates 
yields $\ts \approx (242 \pm 46) \times (f/0.15)$~K (TXS1755+578) and 
$\ts \approx (372 \pm 18) \times (f/1.0)$~K (TXS1850+402), assuming that the \hii\ absorption
towards TXS1755+578 arises towards the radio core. Both absorbers thus appear to have
relatively low $\ts$ values, significantly lower than the typical spin temperatures of high 
redshift DLAs \citep[$\ts \gtrsim 1000$~K;][]{kanekar14}. Low-frequency VLBI imaging of the two 
DLAs will be of much interest to directly estimate the core fraction close to the redshifted 
\hii\ line frequencies.

It should be emphasized that it is possible that the detected \hii\ absorption towards 
TXS1755+578 arises towards one of the four jet components, and not towards the radio core. 
This cannot be ruled out, given the weakness of the \hii\ absorption as well as the relative 
weakness of the core compared to the jet components. Further, the core is likely to have a 
strongly inverted spectrum and is hence likely to be even weaker relatively to the jet 
components at low frequencies. It is hence plausible that the radio and optical sightlines 
are not the same for TXS1755+578 (see below). Caution should hence be exercised while interpreting 
the results towards this source, in the absence of VLBI observations in the redshifted \hii\ line. 

Both DLAs are known to show strong metal-line absorption in their optical spectra, with
multiple absorption components \citep[][]{prochaska98,jorgenson10}. For TXS1850+402,
the redshifts of the two strongest metal-line components are in reasonable agreement 
with the two \hii\ absorption components. These components are likely to be arise in
cold gas, which gives rise to the \hii\ absorption; the other metal-line components
are likely to originate in warmer gas. Conversely, in the case of TXS1755+578, \citet{jorgenson10}
found eight \ci\ absorption components in their Keck-Hires spectrum, as well as \sitwostar\ 
absorption. Since \ci\ absorption is expected to arise in cold gas, it may be surprising that 
the \hii\ absorption profile towards TXS1755+578 shows only a single absorption component,
which is itself offset in velocity from the stronger \ci\ lines. This too suggests that radio 
core in TXS1755+578 may be extremely weak at low frequencies, with the \hii\ absorption
arising towards one of the components in the radio jet. As such, one should not use 
the \hi\ column density determined towards the optical QSO with the \hii\ optical depth 
to estimate the absorber spin temperature.

Finally, the DLA towards TXS1850+402 has a relatively high metallicity, [Zn/H]~$= -0.68 \pm 0.04$
\citep[][]{prochaska98}. Its low $\ts$ value is thus consistent with the anti-correlation 
between metallicity and spin temperature that has been found in DLAs 
\citep[][]{kanekar09c,ellison12,kanekar14}. The DLA towards TXS1755+578 has an even higher 
metallicity, [Zn/H]~$=-0.25 \pm 0.19$ \citep[][Jorgenson et al., in prep.]{jorgenson10}.
While this too is consistent with a large cold gas fraction and hence a low DLA spin 
temperature, the likely difference between the radio and optical sightlines in this 
absorber implies that one cannot test the anti-correlation between metallicity and 
spin temperature here.

In summary, I report the detection of redshifted \hii\ absorption in two DLAs at $z \approx 2$ 
towards TXS1755+578 and TXS1850+402 with a new wide-band GMRT receiver that covers $250-500$~MHz,
with integrated \hii\ optical depths of $\int \tau_{\rm 21cm} {\rm dV} = (0.85 \pm 0.16)$~\kms\ 
(TXS1755+578) and $\int \tau_{\rm 21cm} {\rm dV} = (2.95 \pm 0.14)$~\kms\ (TXS1850+402).
These are only the eighth and ninth detections of \hii\ absorption in DLAs at $z \gtrsim 2$.
For the $z= 1.9888$ DLA towards TXS1850+402, the compact nature of the background source (with
size $\approx 1.4$~mas at 5~GHz) suggests that the low-frequency covering factor is close to unity. 
Combining the \hi\ column density measured from the Lyman-$\alpha$ line with the \hii\ optical depth 
then yields a spin temperature $\ts = (372 \pm 18) \times (f/1.0)$~K. This low $\ts$ value and the 
relatively high DLA metallicity are consistent with the anti-correlation between the two quantities
that has been earlier found in DLAs. In the case of the $z = 1.9698$ absorber towards TXS1755+578,
the weakness of the radio core and the fact that the profiles in the \hii\ and \ci\ lines 
are very different suggest that the radio and optical absorption arise from different sightlines,
and hence, that one should not attempt to combine the two to infer a spin temperature. The 
detection of two new \hii\ absorbers in the commissioning phase of the new wide-band GMRT receivers,
and with only one-third the collecting area of the full GMRT, indicate that both the number of 
detections of \hii\ absorption and our understanding of physical conditions in the neutral gas 
in high-$z$ DLAs are likely to improve significantly in the next few years.

\acknowledgments
It is a pleasure to thank Hanumanth Rao Bandari, Jayaram Chengalur, Yashwant Gupta, Shilpa Dubal, 
Santaji Katore, Navnath Shinde, Rupsingh Vasave, Nilesh Raskar, Deepak Bhong and Manisha 
Jangam for many discussions on the new GMRT receivers and much help with the observations. 
I also thank Regina Jorgenson for providing, in advance of publication, the metallicity of the 
DLA towards TXS1755+578, and Jayaram Chengalur, Maryam Arabsalmani, and an anonymous referee 
for comments on an earlier version of the manuscript. Finally, I thank the staff of the GMRT who 
have made these observations possible. The GMRT is run by the National Centre for Radio 
Astrophysics of the Tata Institute of Fundamental Research. I acknowledge support from the 
Department of Science and Technology, India, via a Ramanujan Fellowship. 

\bibliographystyle{apj}
% \bibliography{ms}

\end{document}